# The Intensities of Cosmic Ray H and He Nuclei at ~250 MeV/nuc Measured by Voyagers 1 and 2 - Using these Intensities to Determine the Solar Modulation Parameter in the Inner Heliosphere and the Heliosheath Over a 40 Year Time Period

W.R. Webber[1], E.C. Stone[2], A.C. Cummings[2], B. Heikkila[3], and N. Lal[3]

1.  New Mexico State University, Astronomy Department, Las Cruces, NM  88003, USA
2.  California Institute of Technology, Space Radiation Laboratory, Pasadena, CA  91125, USA
3.  NASA/Goddard Space Flight Center, Greenbelt, MD  20771, USA

## ABSTRACT

We have determined the solar modulation potential, $\phi$, vs. time that is observed at Voyager 1 and 2 from measurements of the H and He nuclei intensities at a common energy of ~250 MeV/nuc. The H nuclei have a rigidity ~0.7 GV, the He nuclei ~1.4 GV. These measurements cover a 40 year time period, which includes almost 4 cycles of solar 11 year sunspot variations, throughout the inner heliosphere out to the HTS at distances of 95 AU and 85 AU, respectively at V1 and V2, and then beyond in the heliosheath. Inside the HTS the modulation potential vs. time curves at V1 and V2 show a very similar temporal structure to those observed at the Earth. During a later period of maximum solar modulation from ~2000.0 to 2005.0 when V1 and V2 are in the outer heliosphere between ~60-94 AU, the main temporal features of the modulation potential curves at all 3 locations match up with appropriate time delays at V1 and V2 if it is assumed that spatially coherent structures are moving outward past V1 and V2, with outward speeds of up to 700 Km·s$^{-1}$. After 2004.0 V1 and V2 are at latitudes of +35° and -30° respectively, placing lower limits on the latitude extent of these structures. Beyond the HTS in the heliosheath the modulation potential slowly decreases at both spacecraft with only a weak evidence of the unusual modulation minimum observed at the Earth in 2009, for example. A sudden decrease of the modulation potential ~50 MV for both H and He nuclei occurs at V1 just before the heliopause crossing at about 122 AU. This decrease has not yet been observed at V2, which is now at ~113 AU and still observing a modulation potential ~60 MV.

**<u>Introduction</u>**

In their journeys through the heliosphere, Voyagers 1 and 2 have sampled the effects of solar modulation on galactic cosmic rays over a nearly 40 year time period.  This time period covers almost four 11 year solar cycles and includes modulation effects in both the North and South hemispheres to latitudes of $\pm$ 35°.  This includes the inner heliosphere out to the HTS which is at between 84-94 AU and also the heliosheath of thickness ~27 AU, out to a distance at V1 corresponding to the heliopause which was crossed at 122 AU by V1 in 2012 (Stone, et al., 2013).

The study of the changing spectra of H and He nuclei from ~3-600 MeV/nuc throughout this mission can be used to describe this modulation in terms of known, but obviously incomplete, solar modulation models.  During much of the mission, ACR for both H and He nuclei are present and obscure the modulation effects below ~100 MeV/nuc.  We have found that between ~150-350 MeV/nuc we can obtain a very clean and accurate measurement of both of the H and He intensities.  The modulation effects can thus be studied at a common energy 250 $\pm$ 100 MeV/nuc for both H and He nuclei at both V1 and V2.  Since, at this common energy, the mean rigidity of H is 0.71 GV and that of He is ~1.42 GV, the intensity changes of these nuclei will depend on the values of the diffusion coefficient and its rigidity dependence and the other parameters including outward convection that are used to describe the modulation above about 1.0 GV.

The level of solar modulation as deduced from Voyager measurements can also be compared with corresponding measurements at the Earth, in particular the studies of He nuclei by Mewaldt, et al., 2010, 2016 and Lave, et al., 2013, which use both IMP-8 data and ACE data on heavier nuclei at comparable energies to derive a modulation coefficient, as well as neutron monitor measurements at the Earth of this modulation parameter at higher energies using a force field approximation similar to that used in this paper to estimate the incident cosmic ray spectra at the Earth (Usoskin, et al., 2011).

## Voyager Measurements of the H and He Intensities

Figures 1 and 2 show Voyager 1 and 2 measurements of the H and He intensities at 250 $\pm$ 100 MeV/nuc from launch to the present time. This data are 26 day average ratios. V1 moves outward ~3.6 AU/year in the N Hemisphere, passing through the inner heliosphere and reaching the HTS about 27 years after launch at a distance of 94 AU. It then takes about 7.5 years to cross the heliosheath, and V1 has spent the last 4 years beyond the heliopause in local interstellar space.

V2, moving outward at ~3.15 AU/year in the S hemisphere, took 30 years to reach the HTS at a distance of 84 AU and is now in the outer heliosheath, just inside the heliopause, ~9.5 years and 29 AU later. Both spacecraft have been between 30°-35° N-S latitude since about 2004. At 1989 when V2 was still in the ecliptic plane, V1 was at 33° N, a maximum latitudinal difference between the spacecraft.

The H and He intensities shown in Figures 1 and 2 are both at a minimum in 1981 and then increase to a maximum beyond the HP which was crossed in August, 2012, at V1. The ratio of maximum to minimum intensities is ~41 for H nuclei and ~13.5 for He nuclei. There are intensity peaks near the time of launch in 1977 and again in 1987 and again in 1998 at both V1 and V2, almost exactly matching the peaks in the 11 year solar modulation cycle observed at the Earth for He nuclei at a similar energy, e.g., Mewaldt, et al., 2010.

After 2005.0, when V1 crossed the HTS, the H and He intensities at V1, shown in red, continue to increase in an almost linear fashion up to 2012.2 just before V1 crossed the HP. At that time there is a complex jump in both the H and He intensities. After that time the intensities have remained constant and these are assumed to be the LIS values (Stone, et al., 2013).

The intensities at V2, which follows V1 outward in radial distance, are very similar showing intensity maxima at the time of launch and again in 1987 and 1998. At the following solar modulation maximum which occurs between 2001-2005, the H and He intensities exhibit a complex minimum intensity time profile at both spacecraft. The intensities at this time are notably less at V2 than at V1, which is consistent with more solar modulation at V2. This is also consistent with the fact that V2 is now trailing V1 in radial distance by an average of almost 20 AU, along with a radial intensity gradient in the inner heliosphere at that time.

However, from about 2004.5 on the H and He intensities measured at V2 begin to increase rapidly, continuing through the HTS crossing at 84 AU at 2007.7 and then at a slower rate up to 2017, including a dip at 2010.2. Between 2012.0 and 2017.0 the increase is only ~10%.

We believe that these time histories of the intensities at V1 and V2 replicate the inverse of the amount of solar modulation (the value of the modulation potential, $\phi$) observed at V1 and V2 during this overall time period. Note that at 2017.0 when V2 is at 113 AU there is still solar modulation at V2 as evidenced by the fact that the H and He intensities are ~40 and 25%, respectively, lower at V2 than at V1 which is already in the LIS medium.

## <u>The Solar Modulation and the Predictions of the Changing H and He Nuclei Intensities</u>

In this paper we use the spherically symmetric force-field model from Axford and Gleeson (1968) to describe the solar modulation that is observed at any location in the heliosphere. As Gleeson and Axford, 1968, demonstrated this "energy loss" description is equivalent to the results of a spherically symmetric transport description where the particle motion in the heliosphere is described by isotropic diffusion, K, along with an outward flowing solar wind with radial speed V (Km/sec). The motion of particles is described as a diffusion against the solar wind, leading to convection and energy loss. The simple force field solution obtained by Gleeson and Axford, 1968, arises when the diffusion coefficient is ~$\beta$P. In this case a single modulation parameter describes completely the total amount of modulation. This parameter is called $\phi$ (in MV). A potential energy $\Phi = \phi$ Z/A describes the average energy loss of particles (mass A, charge Z) from their energy in interstellar space, so, in fact, $\phi$ is a potential difference. The quantity $\phi$ is sometimes also called the solar modulation parameter. The calculation of the particle flux $J_{mod}$ (E) and any location inside the heliosphere is described by the simple formula:

$$J_{mod} (E, r) = J_{LIS} (E + \phi) \cdot [E (E + 2E_r)]/[(E+\phi) \cdot (E + \phi + 2E_r)]$$

where E is the kinetic energy in MeV/n, $E_r$ is the rest energy also in MeV/n and $J_{LIS}$ $(E + \phi)$ is the interstellar flux, dj/dE, at an energy E+$\phi$, in particles/m²·sr·sec·MeV. Note that if $E_r$ is small, as is the case for electrons, the above equation becomes

$$J_{mod} (E, \phi) = J_{LIS} (E + \phi) \cdot [E^2/(E + \phi)^2]$$

The only temporal variable here is $\phi$, so using this formula with the modulation parameter $\phi$ as a free parameter, we have modulated the input interstellar H and He spectra that have been derived from those measured at V1 (see Cummings, et al., 2016).

This simple formulation is useful because the modulation parameters derived from this "force field" approach are essentially the same as these obtained from calculations using the full spherically symmetric transport model when the diffusion coefficient, K ~$\beta$P (see Gleeson and Axford, 1968). In fact, the value of $\phi$ at any radial point, r, in the heliosphere is just

$$\phi = \int_r^R Vdr/3K$$

where R is the location of the modulation boundary and V and K are the average solar wind speed and the diffusion coefficient respectively.

The values of the modulation parameter that we obtain from the Voyager 1 and 2 data using these calculations may then be directly compared with the studies of the modulation parameter, $\phi$, obtained using ACE data at the Earth at similar energies using a transport model approach, see Lave, et al., 2013, and also to determinations of $\phi$ at higher energies using neutron monitor data response curves and assumed LIS spectra and a force field modulation approach identical to ours (Usoskin, et al., 2011).

Using the formulations in the SACLAY program that we use for this calculation, which is both a LBM for the propagation of cosmic rays in the galaxy (called AMI2), Engelmann, et al., 1990, and a calculation of the solar modulation of this LIS spectrum using the Gleeson and Axford "force Field" approach (which is a sub-routine in this program), we have first calculated the local spectra near the heliosphere that fit the observed H and He spectra observed by Voyager between ~30-300 MeV/nuc (Stone, et al., 2013) where the initial source spectra of both H and He are taken to be (dj/dP) $\sim P^{-2.26}$, where P is rigidity. These source spectra are modified by diffusive and energy loss processes during galactic propagation in the LBM that we use. The interstellar path length ($\lambda \sim \beta/K$) is taken to be $\beta\, P^{-0.45}$ above a lower rigidity which is between 0.316 and 1.00 GV and it is $\sim\beta$ below these rigidities (Webber, 2016). These (average) parameters in the galaxy are obtained from fits to the B/C ratio at both low and high energies from ~10 MeV/nuc to almost 1000 GeV/nuc using Voyager and AMS-2 data on this ratio (Webber and Villa, 2016) as well as fits to the Voyager measured LIS spectra of the H and He nuclei between 3-600 MeV/nuc (Cummings, et al., 2016).

Figure 3 shows the individual H and He nuclei intensities $j_H$ and $j_{He}$ in P/m$^2$·sr· sec·MeV/nuc as a function of $\phi$ in MV that are calculated using the SACLAY program at an energy of 236 MeV/nuc using the LBM model for interstellar propagation and then the "force field" modulation acting upon this spectrum. These calculations are made for 14 separate modulation potentials between $\phi$=0 and 1200 MV. This curve is then fit with a 5$^{th}$ order polynomial and the resulting curve, j vs. $\phi$, is then used to convert the H and He intensities as a function of time measured by V1 and V2 as shown in Figures 1 and 2, into corresponding modulation potential vs. time curves. This conversion is not linear. At a modulation potential of 200 MV a change in potential of $\pm$10 MV produces a change of intensity of $\pm$ 4.3%, at 400 MV, the same potential change of $\pm$ 10 MV produces a $\pm$3.5% change of intensity and at 800 MV the intensity change is $\pm$2.7%. The modulation parameter as a function of time thus obtained using this transformation from the observed V1 and V2 intensities for H nuclei is shown in Figure 4 and for He nuclei in Figure 5.

## A Discussion of the Modulation Parameter $\phi$ vs. Time Curves for V1 and V2 in Figures 4 and 5

At the times of sunspot minimum in 1977, 1987 and 1998, the $\phi$ vs. time curves for both H and He nuclei for V1 and V2 in Figures 4 and 5 exhibit minima in the modulation potential of ~320 MV, 300 MV and ~210 MV, respectively. At times of maximum potential in 1981, 1991 and 2001-2005 the values of $\phi$ increase to 800-900 MV, 700-800 MV and 300-400 MV respectively.

It is to be noted that the values of $\phi$ for both V1 and V2 generally agree to within $\pm$ 10-20 MV (e.g., the intensities are within $\pm$ 5-10%) at times before ~2000.0 in spite of considerable differences in radial distance and also in N and S latitude. One period is observed in 1987 at V2 for about a year in duration when the potentials derived for both H and He nuclei are much lower (>50 MV) at V2 than at V1, implying higher intensities at V2 in the ecliptic plane although V1 was several AU further from the Sun. About 1.5 years later the situation is reversed although V2 was still in the ecliptic plane. This is a time period when significant latitudinal gradients were observed for anomalous cosmic rays, (Webber, Cummings and Stone, 1985). Further comparisons of these 11 year modulation features between V1 and V2, including radial and latitudinal gradients and also a comparison with Earth based studies, will be carried out later in this paper and also in separate articles.

It is also to be noted that, at V1 at the time of the HTS crossing at 2005.0 (see Figures 7-8 which show the modulation potential from 1998 on an expanded time scale), the modulation potential from both H and He nuclei observations suddenly decreased by ~20-30 MV in one 26 day time period. After that the value of $\phi$ continues to decrease rapidly to a value ~180 MV in late 2005.0, a total decrease of $\phi$ by about 100 MV over a time period of a few months which may be associated with the HTS crossing period itself and the region just beyond.

After that the modulation potential at V1 decreased more slowly over about 6.5 years to a value ~50 MV at 2012.3. The modulation potential thus decreased with radial distance at an average rate ~4.25 MV/AU between about 2005.8 and 2012.3, a period ~6.5 years when V1 moves outward about 23 AU, (e.g., most of the heliosheath). After that time both the H and He intensities suddenly increase to this LIS value and the modulation potential becomes ~zero. This unusual modulation of GCR near the HP was first noted in the data shown by Stone, et al., 2013, see also Webber and McDonald, 2013, and has been discussed in some detail by Webber and Quenby, 2015.

Meanwhile at V2, which was still inside the HTS, the modulation potential also began to decrease rapidly from ~420 MV for H and ~330 MV for He nuclei at about 2004.5 to values of ~140 MV for both nuclei at the time of the V2 HTS crossing ~3 years later at 2007.7. A similar decrease of modulation potential at the Earth, in this case from ~800 MV to ~420 MV during this same time period, was reported by Lave, et al., 2013 and was based on C and heavier nuclei data. These decreases in modulation potential in this time period from about 2004.5 to 2007.7, observed at both the Earth and V2, are probably related to the onset of the next 11 year solar sunspot cycle which eventually reached its modulation minimum ~260 MV for He and C nuclei in late 2009 at the Earth (Mewaldt, et al., 2010; Lave, et al., 2013).

There was no specific decrease in $\phi$ as V2 crossed the HTS, however, as was the case for V1, but beyond the HTS the modulation parameter at V2 continued to decrease to a value ~70 MV at about 2011.5 when V2 was at 96 AU, about 11 AU beyond the HTS. After 2011.5, the rate of decrease became almost zero. In the 4 year time period from 2007.7 to 2011.5 when the potential decreased from 120 to 78 MV (for H) the average rate of decrease of $\phi$ at V2 was about 4.5 MV/AU of outward travel, about the same as at V1 during a corresponding time period just beyond the HTS. During the following time period at V2 from 2011.5 to 2017.0, the modulation potential remained between about 50-70 MV. In the time period after 2011.5 V2 appears to be in a "different" region of the heliosheath (e.g., Burlaga, Ness and Richardson, 2014; Intriligator, et al., 2015).

Overall the solar modulation features at both V1 and V2 (and also at the Earth), are very similar inside the HTS and show a dominant 11 year solar modulation cycle. Beyond the HTS in the heliosheath, however, the modulation features are different. In particular, the 11 year solar cycle variation that is so dominant in the inner heliosphere seems to be muted in the heliosheath at V2. At V1 there is a decreasing modulation potential in the heliosheath as the modulation decreases from 180 MV just beyond the HTS to ~50 MV just inside the HP.

## **Possible North-South Differences in the Modulation Potential**

During the time period from launch to about 2000, a period ~23 years, the difference in modulation potential at V1 and V2 for both H and He nuclei was small, generally less than 10-20 MV equivalent to intensity differences of less than $\pm$ 10%.

In 1987, over a one year time period and at a time of minimum modulation, the potentials derived at V2, which is near the ecliptic plane, from both the H and He nuclei were significantly less that those at V1, which is at ~33° N. The modulation potential differences of up to 80 MV seen at this time correspond to H and He intensity differences between V1 and V2 of over 40% with the intensity being higher at V2. V1 in 1987 was at ~33° N and remained at this latitude while V2 was at 0° S and remained at this latitude until 1989. The two spacecraft were separated by ~5 AU in radius from the sun.

We have examined this time period in detail using the maximum latitude of the HCS as determined by the WSO. These current sheets in the inner heliosphere could be considered a departure from the spherical symmetry of the Gleeson and Axford model used to calculate the potential. Figure 5A shows potential vs. time curves for H nuclei from V1 (red points) and V2 (blue points) for the time period from 1986.0 to 1990.0. Note that the potential difference between V1 and V2 (shown as a shaded region) increases from ~10-20 MV at 1986.0 to a max ~80 MV at about 1987.5, when the intensity at V2 in the ecliptic plane is also a maximum. Then the potential difference rapidly decreases to zero in early 1988. At this same time, also, the maximum latitude of the HCS has reached ~30°, which is, in fact, the latitude of V1. Later, as the maximum latitude of the HCS rapidly increases to ~80°, signaling an impeding change of solar magnetic polarity from negative to positive, the potential difference reverses and V1 now has the lowest potential after 1988.7. Between 1986.0 and 1988.7, both the H and He components display a difference of modulation potential consistent with a negative latitude gradient of up to ~1.5% per degree at 1987.5.

## The Time Period Between 2001-2005 and V1 Crossing the HTS at 2005.0

Between 2001-2005 there is a double peak variation of the modulation potential of both H and He nuclei lasting for over 3 years. This time period is just after the change over from positive to negative polarity of the Suns magnetic field (see Figures 7 and 8).

The 1st peak at V2 reaches a maximum at 2001.8 with a modulation potential ~350 MV for H and ~300 MV for He. The second peak is at about 2004.3 with slightly smaller modulation potentials.

A similar pattern of modulation potential (intensity) changes has been reported at the Earth by Lave, et al., 2013, using ACE data for heavier nuclei such as Carbon. She observes peaks in modulation potential of ~1220 MV at 2001.0 and a second peak with $\phi = 1100$ MV at 2003.84, with a minimum in between. The second peaks that are observed at the Earth and V2 in this sequence could be related to the 2003 Halloween event at the Earth. Wang and Richardson, 2009, estimate a travel time of this event to V2, then at a radial distance of 72 AU, of between 0.5-0.6 year. This corresponds to an average speed of about 660 km·s$^{-1}$ to 72 AU.

The onset of the modulation from minimum to maximum at the Earth starts at about 1999.6 when the modulation potential is 510 MV, and is followed by a modulation potential increase of ~650 MV occurring in two steps over a time period less than 1 year. The onset curves of the modulation potential increase at V2 and V1 relative to those at the Earth are shown in an expanded time scale in Figure 6. The time delay between the Earth and V2 for both step 1 and step 2 of the increase in potential is about 0.5 year corresponding to average speeds of ~650 km·s$^{-1}$. During the same time period, the characteristic shape of the modulation potential curves with time remains similar at each location. The spatial structural features related to these intensity changes must therefore have been similar up to at least $\pm$ 35° latitude as they move outward with a time delay of ~0.5 year to V2.

### The Modulation Potential in the Heliosheath

After the HTS crossing (+0.5 year) the modulation potentials of H and He nuclei decrease at V1 during the time period from 2005.5 to 2012.3 in a manner consistent with a more or less linear decrease with an average value ~4.25 MV/AU representing a total modulation potential decrease ~100 MV (see Figures 7 and 8 for an expanded time view of these variations). Near the HP itself V1 observes a sudden decrease of both the H and He modulation potentials beginning at ~2012.35, about 1 AU in front of the final decrease. This decrease is from a value ~50 MV potential to a value ~zero. The thickness of 1 AU deduced for this feature may be related to the thickness for thE outer region comprising the HP.

At V2 the decrease in the modulation potential in the heliosheath between the HTS crossing at 2007.7 and about 2011.5, covering a radial distance ~11 AU beyond the HTS and including a "bump" at 2010.2, is about ~4.5 MV/AU, very similar in magnitude to that at V1 in a corresponding region beyond the HTS.

The value of the modulation potential at the time of the HTS crossing for V2 was ~120 MV for H and ~130 MV for He nuclei. After 2011.5, for other 5.5 years at least, the potential at V2 remains at between 50-70 MV for both H and He, over an additional radial distance of 18 AU. The potential observed at V2 at 2017.0 is roughly the same modulation potential that was observed just before the final drop to ~zero potential at V1.

**Summary and Conclusions**

In this paper we use measurements of the cosmic ray H and He intensities as a function of time at V1 and V2 to determine the total solar modulation and the modulation potential throughout the heliosphere for a 40 year time period extending over parts of 4 solar 11 year cycles. These intensities can be determined to an accuracy ~a few % and these calculations have shown that these intensities are a sensitive indicator of the total amount of modulation in the heliosphere at rigidities ~1.0 GV and larger. These intensities are measured between 150-350 MeV/nuc for both H and He. At an average energy of ~250 MeV/nuc the rigidity of protons is ~0.71 GV and He nuclei is 1.42 GV. These energies are well above those of the ACR that are present in the heliosheath.

This study divides into two time periods. The first period is when V1 and V2 are inside the HTS and the solar 11 year cycle is dominant. The second is when V1 and V2 are outside the HTS, in the heliosheath, where the H and He intensities gradually increase with radius corresponding to a decreasing modulation potential.

**a) The Heliosphere Inside the HTS**

Voyagers 1 and 2 each have spent nearly 30 years in the region inside the HTS, covering 3 periods of minimum modulation and 3 periods of maximum modulation. The resulting profiles of modulation potential vs. time for both H and He nuclei are very similar to the modulation potential profiles (or equivalently intensity-time profiles) that are observed at the Earth for He nuclei at similar energies by Mewaldt, et al., 2010 and also by Lave, et al., 2013, using both IMP and ACE data on C and heavier nuclei; they are also very similar to the modulation potential vs. time profiles obtained at higher energies using NM data, (Usoskin, et al., 2011). These modulation potential vs. time profiles show a distinctive 22 year periodicity at V1 and V2, similar to those at the Earth, as well as 11 year profiles of maximum and minimum modulation potential.

With regard to the two periods of minimum modulation that are observed in 1987 and 1998-99 by V1 and V2 we note that, at the minimum in 1987 there is a very narrow peak in the $\phi$ vs. time curves (at V2) whereas in the following modulation minimum in 1998-99, the $\phi$ vs. time profile is much broader in time at both spacecraft with an almost exponential decrease of the modulation potential from its maximum to minimum value over a time span ~6 years, between 1992 and 1998. The modulation minimum in 1987 occurred at the end of a negative solar magnetic polarity period whereas the minimum in 1998 was at the end of a positive polarity period.

With regard to the three time periods of maximum modulation observed at V1 and V2, we note that from 1981-1984 and again ~22 years later from 2001-2005, the modulation maxima last for several years and consist of a double peak in the modulation potential. The "in between" modulation maximum potential period in 1990-1991 lasts just a few months centered around the times of maximum potentials at 1990.3 and 1991.3 at both V1 and V2. All three of the modulation maxima are double peaked. These periods of maximum modulation occurred just after the change-over in magnetic polarity from negative to positive polarity in 1981-84 and 2001-2005 and the opposite polarity change-over in 1990-1991.

The relative modulation potential that is determined at V1 and V2 may be compared throughout the entire time period when V1 and V2 are inside the HTS from launch to the time they cross the HTS. Throughout most of the time period from launch up to about 2000 (a period of 23 years) the modulation potential at V1 and V2 for H and He nuclei individually are the same at both spacecraft within $\pm$ 10-20 MV implying intensity differences of less than $\pm10\%$. During this time the radial separation of these two spacecraft is increasing by ~0.5 AU/year from about 6 AU at a time of the minimum modulation in 1987 to about 19 AU in 2000. The latitudinal difference between the two spacecraft increases rapidly at first, to ~33° in 1987 as V1 leaves the ecliptic plane. V2 remains near the ecliptic plane during this early time period. Then, starting in 1989 as V2 moves to higher latitude, the latitude difference between the spacecraft rapidly becomes smaller reaching ~18° by 2000 and then decreasing to a difference of less than ~5° at the times of V1 and V2 HTS crossings at 2005.0 and 2007.7, respectively, when both spacecraft are at latitudes ~30-35°.

A comparison of the V1 and V2 modulation potentials in the period before 2000 seems to imply a pattern of small latitudinal ($\leq$ 1% per degree) and radial ($\leq$ 1% per AU) intensity gradients; with non-spherically symmetric structural features, such as current sheets, playing a role in the larger potential differences that are seen between V1 and V2 in 1986-1989.

After 2000, as a result of an increasing modulation due to the solar cycle 11 year variation, a significant modulation potential difference is observed between V1 and V2. This could be explained as primarily a radial intensity gradient ~5%/AU between V1 and V2 and the fact that by 2005.0 when V1 crosses the HTS, the radial separation of the two spacecraft is almost ~20 AU (V1 is at ~35° N, V2 at ~30° S) at this time.

In summary, the above observations suggest that, (1) in the time period from 1977 to 2000 covering parts of three ~11 year solar modulation cycles, the radial and latitudinal gradients between V1 and V2 were, in general, relatively small; and (2) during the maximum modulation time period from 2001.0-2005.0, when V1 and V2 are beyond 70 AU, the radial gradients between V1 and V2 were larger (~5%/AU). At about 2000.0, the solar magnetic polarity changed from positive to negative.

These Voyager observations will provide further important tests of latitude related terms in the modulation calculations made using the full transport models, including the effects of drifts and latitude dependent effects in the diffusion tensor (e.g., see Potgieter, et al., 2013).

The observed similarities in the temporal variation of the modulation potential at V1 and V2 and at the Earth during the modulation maxima time period from 2001.0 to 2005.0 up to the time when V1 crossed the HTS are consistent with two spatially coherent modulation structures moving rapidly outward from the Sun to V1 and V2, which are at average distances of ~71 and 89 AU, respectively, and latitudes + 35° and -30°, respectively. These two structures produced most of the total modulation potential change measured at the Earth, V2 and V1 during the onset of this solar 11 year modulation cycle in 2000 and 2001. This total onset time took only ~1 year at all locations. These individual onset structures reach V2 with a time delay between ~0.5 year, corresponding to average propagation speeds 600-700 Km·s⁻¹ in the heliosphere out to ~70 AU.

Overall, the concept of a simple modulation potential as originally described by Gleeson and Axford, 1968, provides a reasonable description of the general features of the 11 year solar cycle variations above ~1 GV that are observed for H and He nuclei at V1 and V2 throughout the heliosphere up to latitudes of $\pm$ 30° out to the HTS. Departures from a spherically symmetric "force field" picture may be due more to structural features such as current sheets within the inner heliosphere (inside the HTS) rather than systematic radial and latitudinal intensity gradients in the time period up to 2000, at least at rigidities ~1 GV and above. Between 2000 and 2005 the modulation potential differences between V1 and V2 become consistent with a larger radial intensity gradient, along with a zero or a small latitudinal gradient up to ~35° if the latitude gradients are indeed symmetric between the N and S hemispheres.

### b) The Heliosphere Outside the HTS

Beyond the HTS in the heliosheath, significant modulation still exists at ~1 GV and above; however, the imprint of the solar 11 year variation on this modulation potential is not prominent in this region. A total modulation potential between 130 and 180 MV or about 0.3 to 0.5 of the total modulation potential ~300-400 MV observed at the Earth during periods of minimum solar modulation, is observed in this heliosheath region just after both V1 and V2 enter the heliosheath at different times in the 11 year modulation cycle. The more or less linear decrease in modulation potential from a point just outside the HTS to a point just inside the HP at V1 may provide clues as to the origin of the outer heliosphere modulation. Other clues include the sudden decrease in modulation potential ~50 MV seen at V1 which occurs in two steps just before the final crossing of the HP on August 25[th], 2012.

An average decrease in modulation potential ~4-5 MV/AU is observed at V1 over a distance ~24 AU beyond the HTS and also at V2 over a distance ~11 AU directly beyond the HTS. This rate of decrease would be consistent with a force field modulation picture e.g., where $\phi = \int V dV / 3K$ in MV/AU where V~100 km s⁻¹ and K ~1.0 x 10²² cm²·s⁻¹ in the heliosheath.

The features noted above combine to lead to two important conclusions of this paper: 1) There is, on average, a simple decrease in modulation potential as a function of time or radial distance from the Sun at V1 beyond the HTS and; 2) There is a very muted presence of the 11 year solar modulation effects in this heliosheath region of ~25 AU in radial extent at V2; for example, the end of 2009 super-low modulation potential followed by a rapid increase in the modulation potential in 2010 that was observed at the Earth (Mewaldt, et al., 2011) was not seen at V1 and only weakly seen at V2.

**Acknowledgments:**  The authors wish to thank JPL for their unfailing support, both moral and financial, for 40 years since the launch of V1 and V2 in August, 1977.

# FIGURE CAPTIONS

**Figure 1:** The H nuclei intensities observed by V1 and V2 in a common energy/nuc interval 250 $\pm$100 MeV/nuc. Data are 26 day averages from launch in 1977 to 2017. At V1 the H intensity increases from a minimum value in particles/m²·sr·MeV/n of ~3.2 E-01 in 1981, to an interstellar value ~13.5 after the HP crossing on August 25, 2012 for a total factor ~42.

**Figure 2:** Same as Figure 1 but for He nuclei. For He these minimum and maximum values are 0.8 E-02 in 1981 and 1.06 after August 25$^{th}$ for a total factor ~13.5.

**Figure 3:** The calculated variation of the H and He intensities at 236 MeV/nuc as a function of modulation potential using the force field equation. These calculated H and He nuclei intensities vs. modulation potential are used to translate the H and He intensity vs. time measurements shown in Figures 1 and 2 into the V1 and V2 modulation potential vs. time curves for H and He nuclei shown in Figures 4 and 5.

**Figure 4:** The modulation potential vs. time as derived from the H nuclei intensities observed at V1 and V2 from launch in 1977 to the present time. The shaded time periods in this and later Figures include the time periods when the solar magnetic polarity changes.

**Figure 5:** Same as Figure 4 but for the modulation potential derived from He nuclei intensity changes in the heliosphere inside the HTS and in the heliosheath beyond the HTS.

**Figure 5A:** Modulation potential differences observed at V1 and V2 from 1986.0 to 1990.0. The heliospheric current sheet maximum N latitude from the WSO is also shown. Light shaded region is where $V2_{MP} > V1_{MP}$, dark shaded regions is where the opposite is true.

**Figure 6:** Modulation potential changes for H nuclei occurring at the Earth, V2 and V1 during the time period from 1999.0 to 2002.0 (26 day averages). Events 1 and 2, when the modulation potential increased from 510 to 845 MV and then from 845 to 1160 MV at the Earth are shown along with the estimated time delay, (1) and (2) to V2 as determined from the midpoints of each potential increase at the Earth and V2.

**Figure 7:** The modulation potential vs. time on an enlarged time scale for the time period 1998 to the end of data at V1 and V2 as derived from the H nuclei intensities. Note the more or less uniform decrease in modulation potential after the HTS crossing at ~2005.0 at V1 and after ~2007.7 at V2 (for ~5 years) and also the final decrease of 50-60MV starting just before the HP crossing at V1.

**Figure 8:** Same as Figure 7 but for the modulation potential as derived from He nuclei at V1 and V2.

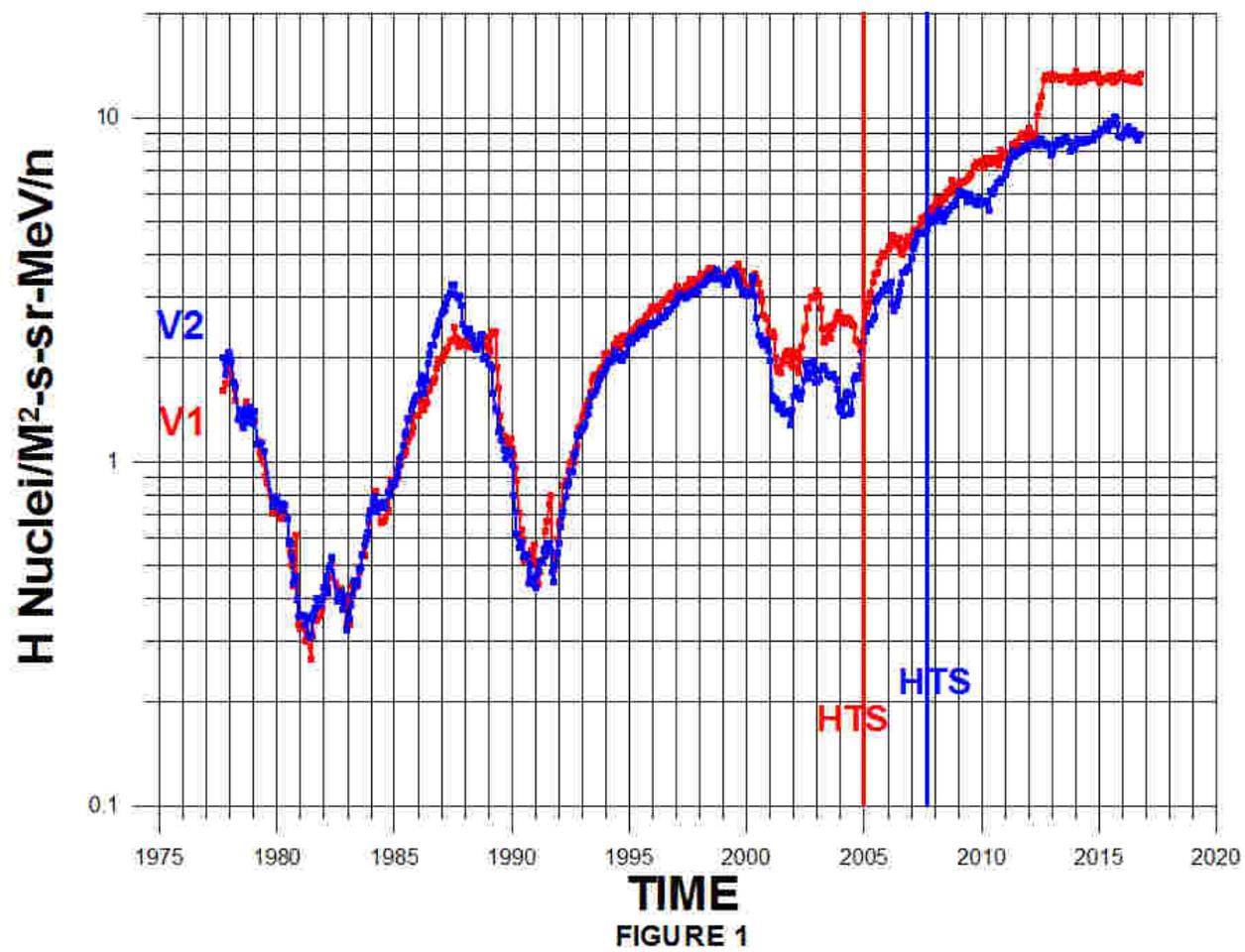

**FIGURE 1**

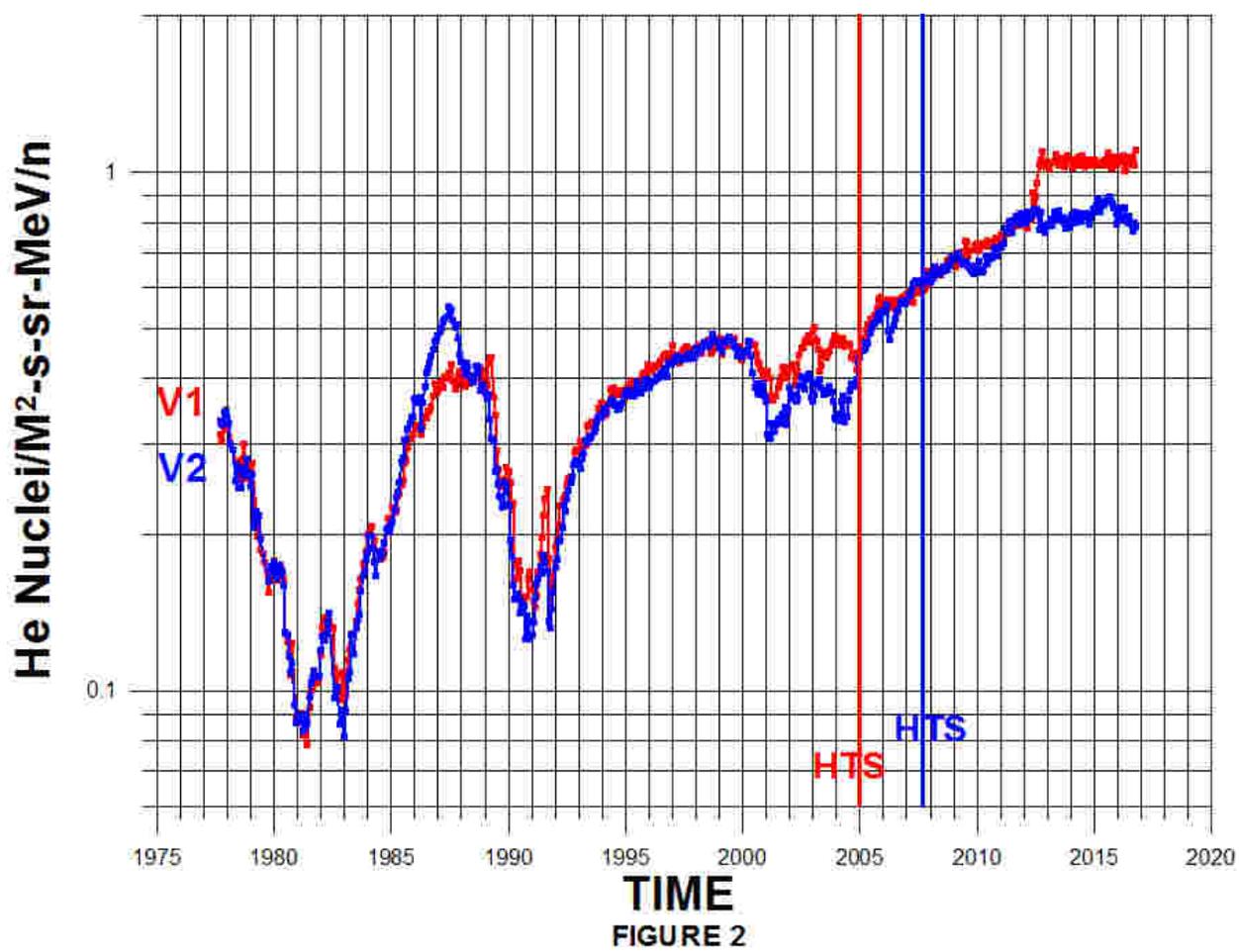

**FIGURE 2**

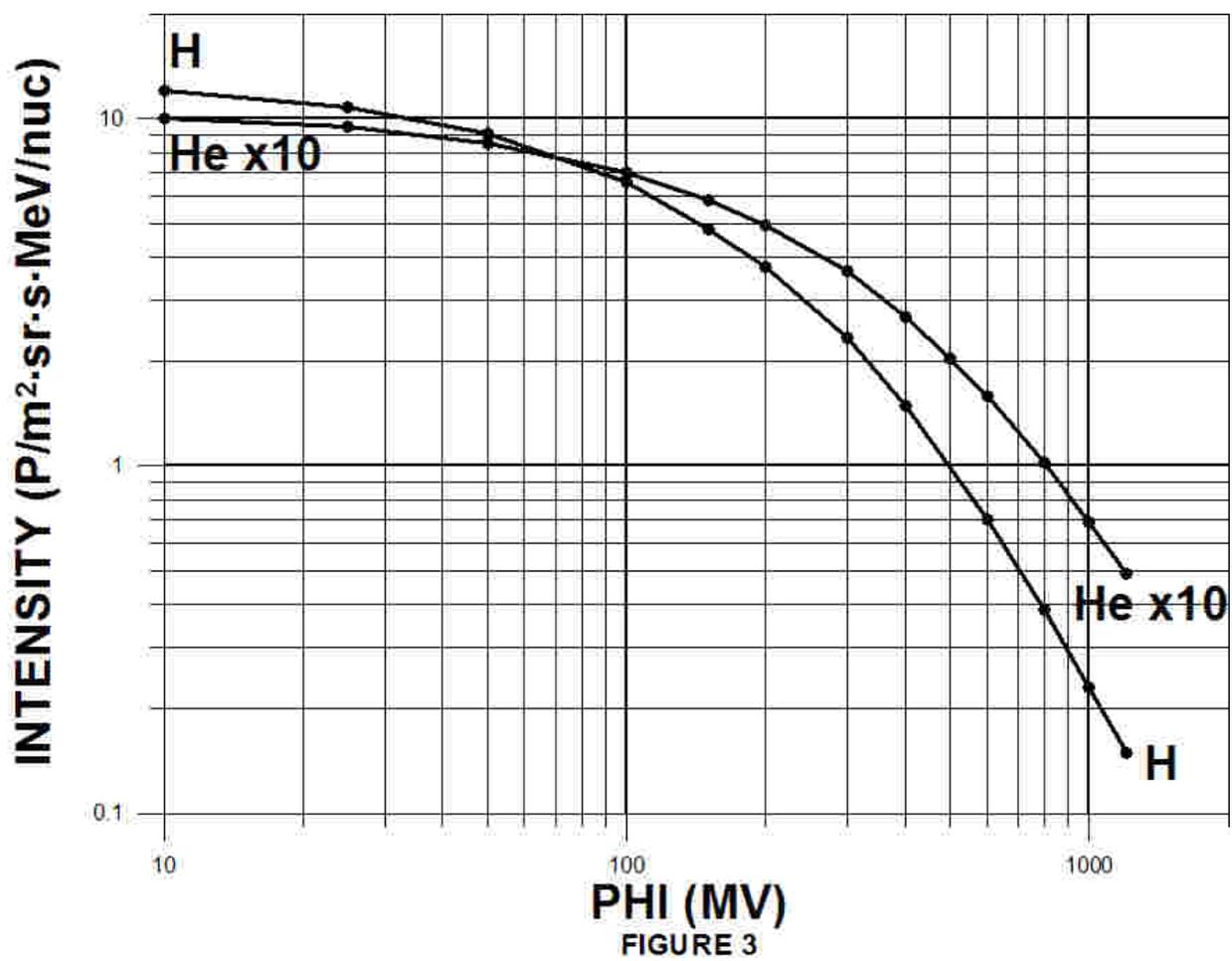

**FIGURE 3**

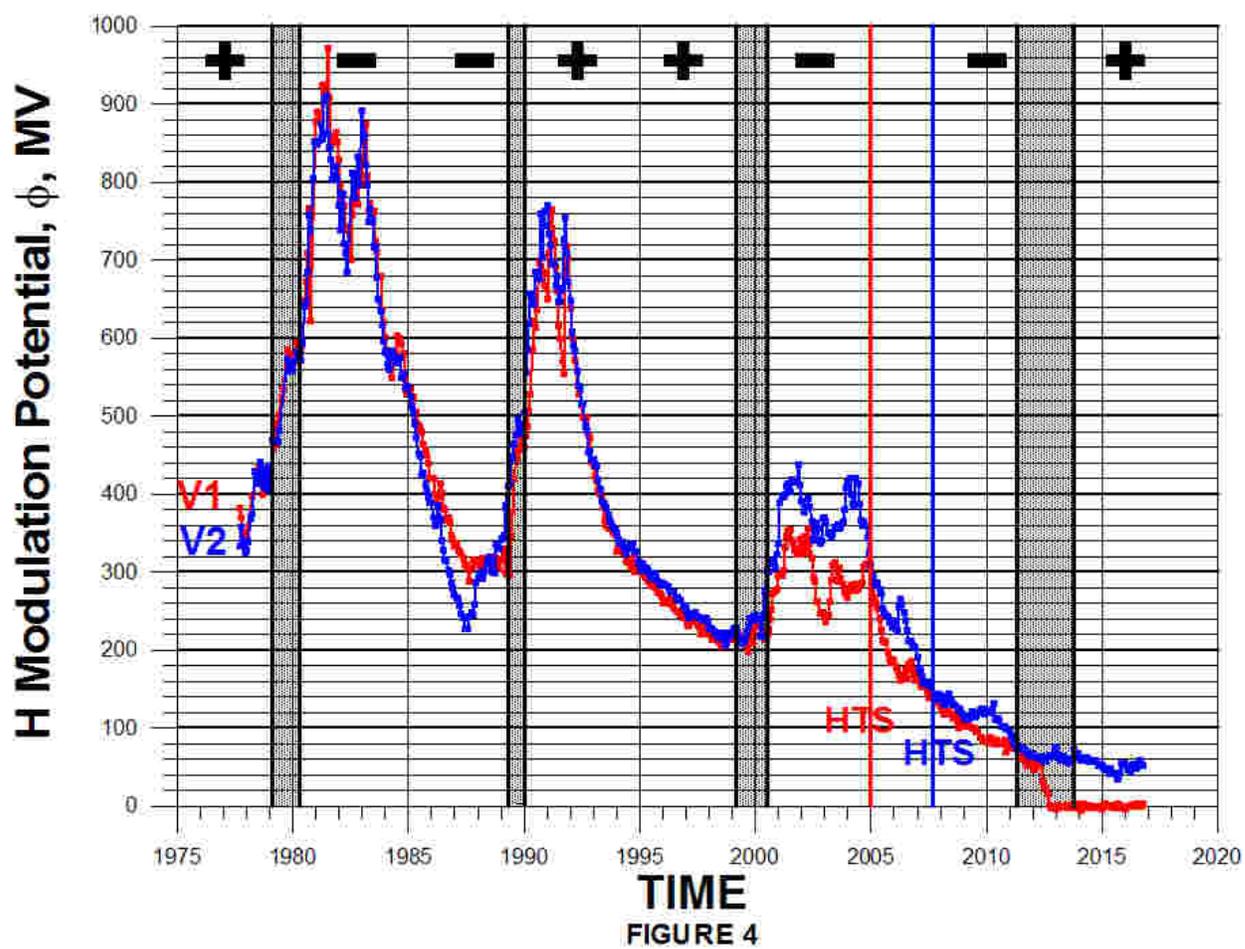

FIGURE 4

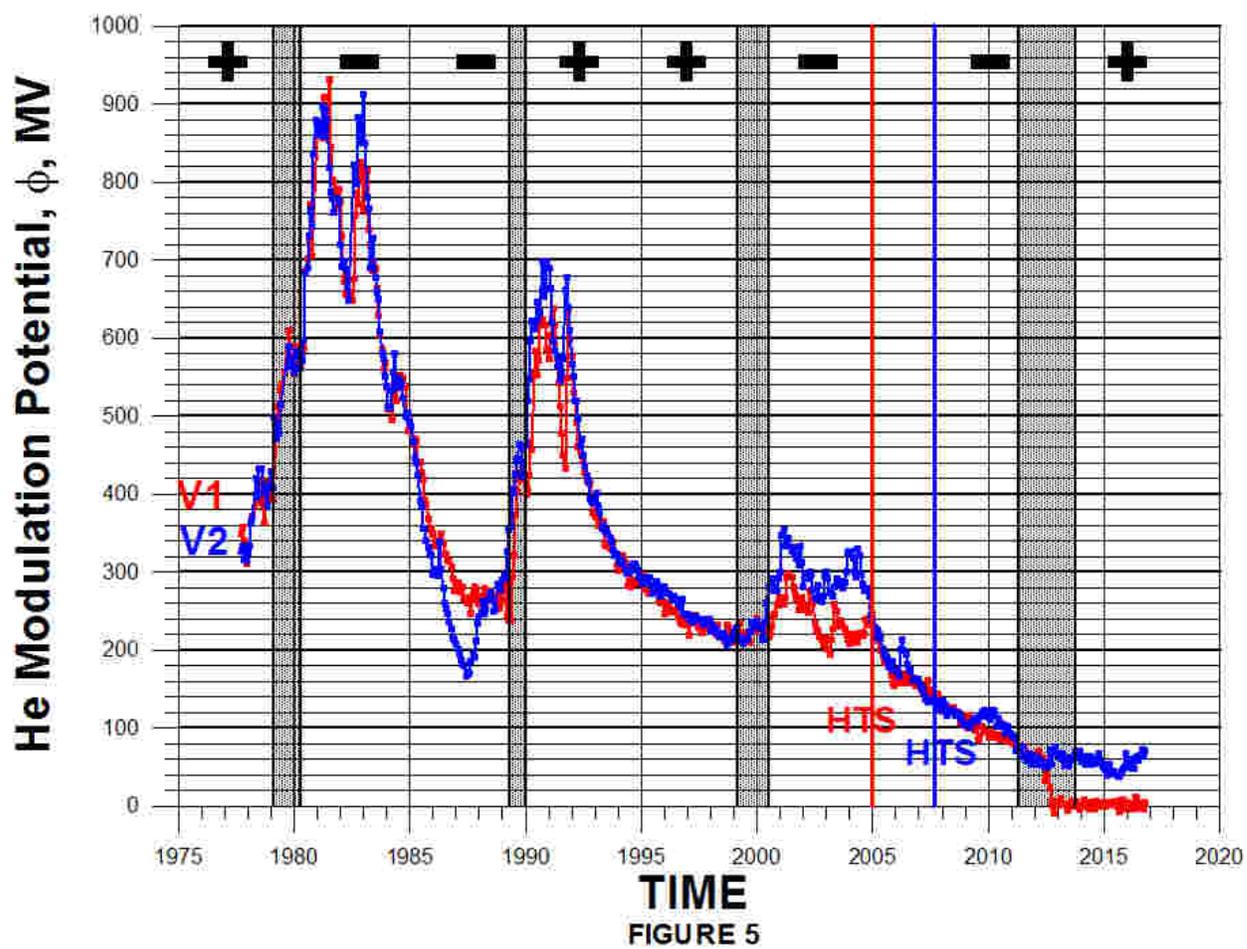

FIGURE 5

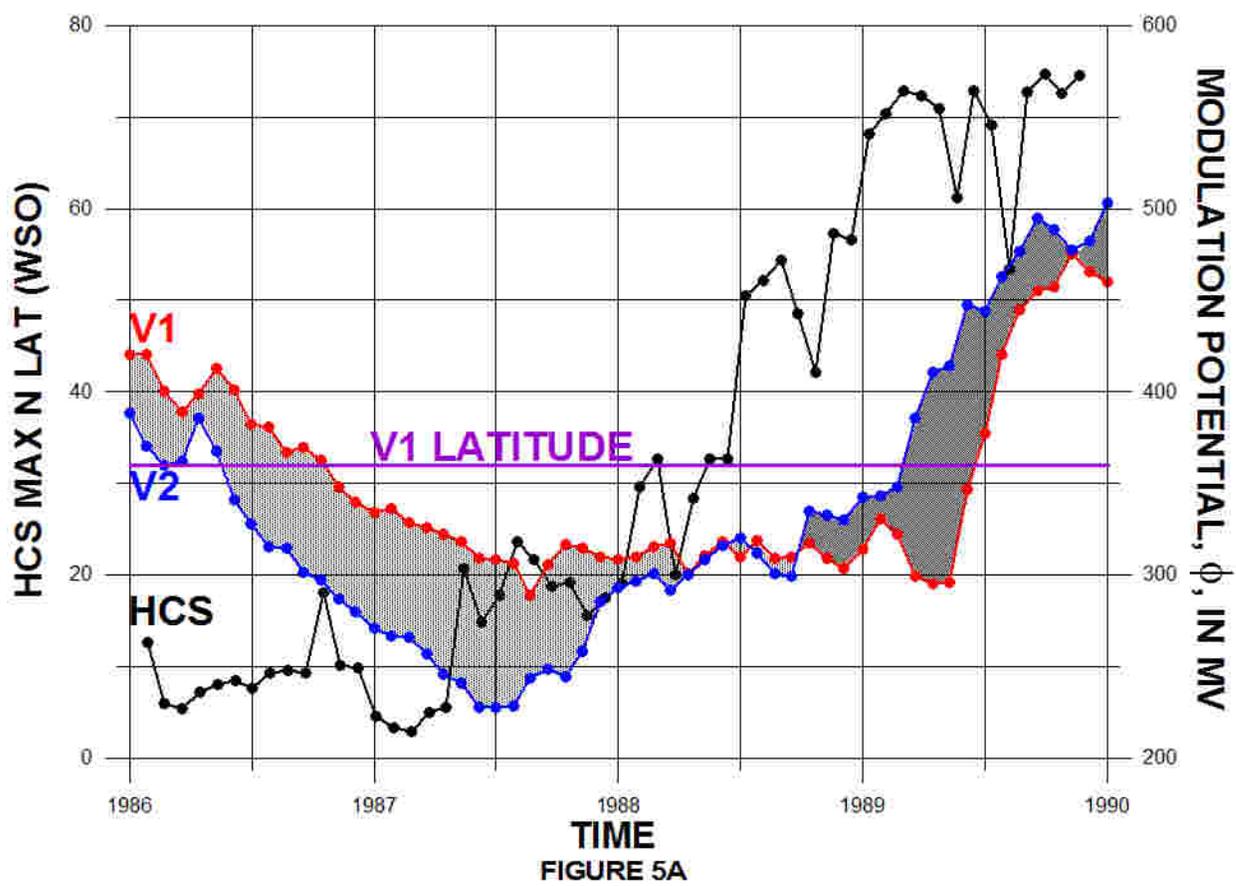

**FIGURE 5A**

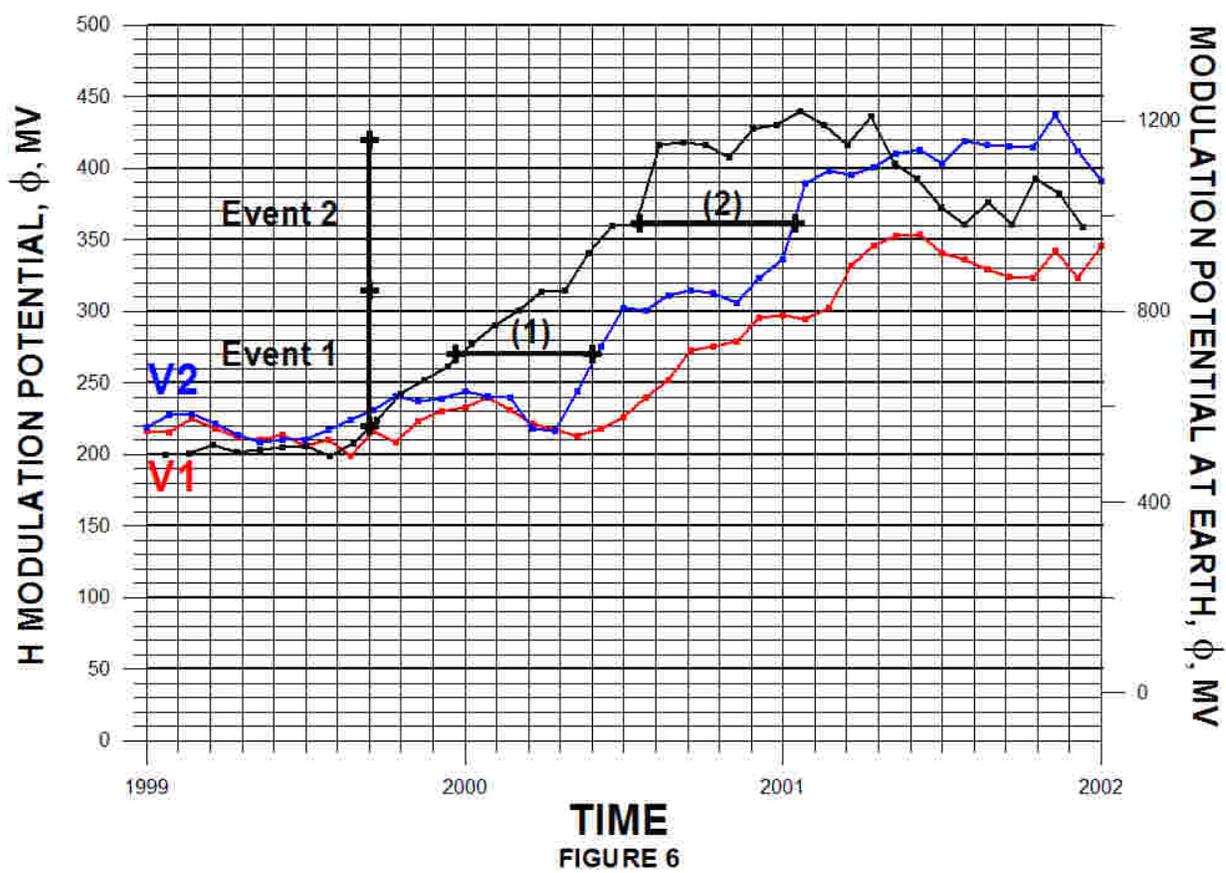

**TIME**
**FIGURE 6**

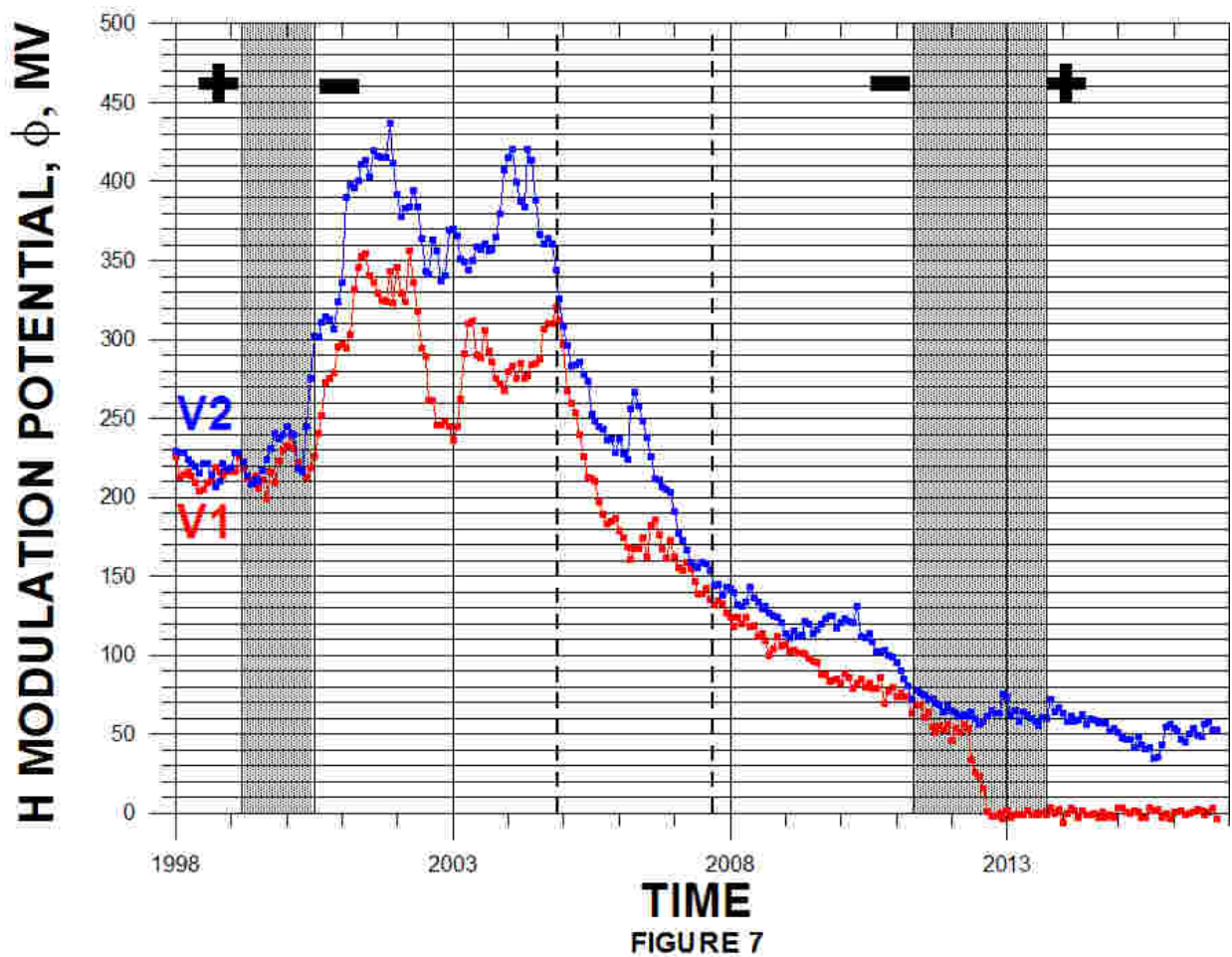

FIGURE 7

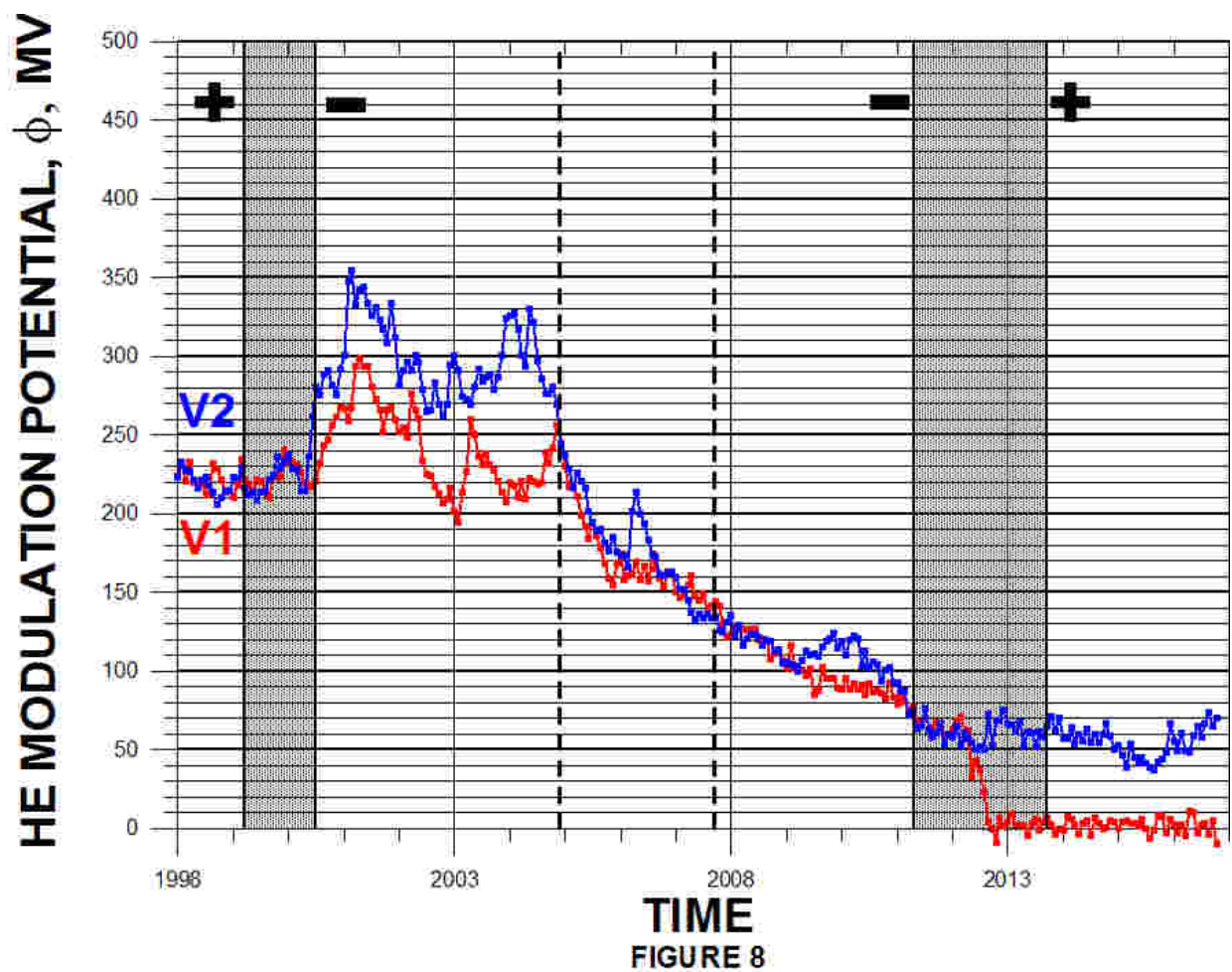

FIGURE 8